\RequirePackage[2018-12-01]{latexrelease}
\documentclass[showpacs,showkeys,11pt,
preprint,preprintnumbers,nofootinbib,
groupedaddress,superscriptaddress,amsmath,amssymb]{revtex4}
\usepackage{amsfonts}
\usepackage{amsmath}
\usepackage{graphicx,subfigure}
\usepackage{caption}
\usepackage[export]{adjustbox}
\usepackage{epsfig}
\usepackage{url}
\usepackage{multirow}
\usepackage{feynmp}

\newcommand {\be}{\begin{equation}}
\newcommand {\ee}{\end{equation}}
\newcommand {\ea}{\end{eqnarray}}

\begin{document}

    \title{Effect of Polarized Colliding Beam on Higgs Boson production at the Lepton Collider}

\author{Ijaz Ahmed}
\email{Ijaz.ahmed@fuuast.edu.pk}
\author{Mashal Shakar}
\email{mashal.mashal111@gmail.com}
\affiliation{Federal Urdu University of Arts, Science and Technology, Islamabad Pakistan}

\author{M. U. Ashraf}
\email{muhammad.ashraf@uclouvain.be}
\affiliation{Centre for Cosmology, Particle Physics and Phenomenology, Université catholique de Louvain, Belgium}
\author{Jamil Muhammad}
\email{mjamil@konkuk.ac.kr}
\affiliation{Sang-Ho College, and Department of Physics, Konkuk University, Seoul 05029, South Korea}
\author{Taimoor~Khurshid}
\email{taimoor.khurshid@iiu.edu.pk}
\affiliation{International Islamic University, H-10, Islamabad Pakistan}

\begin{abstract}
Different production processes involving the Higgs boson, such as annihilation and W/Z boson fusion, will be observed in the International Linear Collider (ILC). The ILC operates at a center-of-mass (CM) energy of $\sqrt{s}$ = 200-1000 GeV. The study reveals that the production cross-section can either be enhanced or reduced depending on the CM energy and the specific combination used, which has implications for selecting appropriate production processes. Additionally, this investigation highlights that by polarizing beams, the number of measurable observables increases. These observables, such as left$~$right asymmetry, detailed effective polarization, and adequate effective luminosity, are crucial to ascertain contemporary physical parameters in physics models absurdly the Standard Model (SM).


\end{abstract}
\maketitle
\section{Introduction}
The degree of alignment of particle's spin angular momentum with its direction of motion is basically spin polarization. The electron beam is considered as quantum system with mixed states of spin orientations. The polarization happens when their spin orientations are divided into contrasting groups. When one of the population of spin orientation is preferred over the other orientation population, the beam is said to be polarized. On the contrary, when all the spins of population of particles are having same orientation, then the beam is totally polarized. If major portion of spins has similar orientation, the beam is considered as partially polarized and in the case when both the populations of orientation are equally distributed, the beam is said to be unpolarized. The polarization can be described as a bunch of particles, having $\lambda =$ -1/2 or $\lambda =$ +1/2, which is helicity for right and left-handed particles respectively~\cite{lab1}. The electron beam can be polarized in two ways; transverse and longitudinal. In longitudinally polarized beam, the spin vectors are symmetrical with the direction of their movement, on the other hand, the transversely polarized beam is one with spin vectors are orthogonal to their traveling direction~\cite{lab2}. A preprint of this study has previously been published as arxiv of Cornell University \cite{prep}.

\section{Exploitation of Polarized beams at Colliders}
In the presence of both fields i.e. electric and magnetic, the spin motion of particles is altered by two depolarization effects:
\begin{itemize}
\item \textbf{Spin Precession (Classical):} 
It is governed by Thomas-Bargmann-Michel-Telegdi (T-BMT) equation. This effect is observed to be dominant at lower energies. Depolarization of about 0.17\% is predicted at International Linear Collider (ILC) and 0.10\% at the proposed Compact Linear Collider (CLIC) for 100\% polarized beams~\cite{lab3}.
\item \textbf{Spin-Flip Processes (Quantum mechanical):} 
Spin-flip processes is also called Sokolov Ternov (S-T) effect through synchrotron-radiation-emission due to which the depolarization effect increases with the increase in energy. The depolarization of about 0.05\% is observed at the ILC and 3.4\% at the CLIC, due to this effect for 100\% polarized beams~\cite{lab3}.
\end{itemize}
About 0.2\% of total depolarization is predicted at the ILC during each bunch crossing[Ref??]. Even more, a higher degree of depolarization is observed at circular collider, the Large Electron-Positron Collider (LEP). Electron polarization was first implemented in Stanford linear collider at the SLAC, which could accelerate the beam of electrons upto energy of 50 GeV and about 80\% to 90\% of polarization of electrons was achieved~\cite{lab4}. The photocathode technology was strained as a source of generating polarized electrons. Undulator radiation is the primary method employed for the generation of polarized positron ($e^+$) while the other two being Compton back-scattering and bremsstrahlung for polarized $e^-$ beam. In order to completely incorporate the polarized beams in accelerators, it is important to precisely measure degree of polarization and hence highly precise polarimetry is necessary~\cite{lab5}. In SLAC, the precision of about 0.5\% in $\frac{\Delta P_{e^-}}{P_{e^-}}$ is already attained. But at the ILC, goal was to make it more precise to the value of $\leq$ 0.25 $\%$ \cite{lab1}.

\subsection{Longitudinal polarization of beams at the Linear Collider}
 Polarization of beams is useful at energy $\sqrt{s}$ = 500 GeV which is considered as the primary stage of ILC, and even benefits at 1000 GeV and at CLIC, which is another linear collider working between range of 1 TeV to 3 TeV. Exploring outcomes by longitudinally polarized \textit{e}-beam undertaking EW process and also the spin-related properties of particles which initiated by the helicity values expelled the electron 80\% degree which are longitudinally polarized, eventually. Most recent results explore larger values, i.e upto 90\% ~\cite{lab6}.  Homogeneously the positron polarized beam 60\%, with no luminosity loss. A higher polarization value of about 75\% can be achieved which will cause loss in luminosity.
 
 
 \section{The Production of Cross-section via Polarized $e^- e^+$ Beams}
 
The longitudinally polarized beams cross section in an $e^- e^+$ collider can be written as:

\begin{multline}
\sigma_{P_{e^+},P_{e^-}} = \frac{1}{4} \{( 1 + P_{e^+} ) ( 1 + P_{e^-}) \sigma_{RR} + ( 1 - P_{e^+} ) ( 1 - P_{e^-} ) \sigma_{LL} \\
+ ( 1 + P_{e^+} )  ( 1 - P_{e^-} ) \sigma_{RL} + ( 1 - P_{e^+} ) ( 1 + P_{e^-} ) \sigma_{LR}\}
\label{eq:1}
\end{multline}

In experiments involving $e^- e^+$ collisions, two primary processes are observed to be prominent or dominant:

\begin{enumerate}
    \item Scattering processes, including t-channel and u-channel interactions, as well as vector boson fusions.
    \item Annihilation processes, specifically the s-channel process known as Higgsstrahlung, which involves the production of the Higgs boson along with a neutral Z boson.
\end{enumerate}
In the annihilation process of $e^- e^+$ collisions, the helicities of the particles are coupled with their spin, and the interchange of these particles occurs in a direct channel. According to the Standard Model (SM), annihilated vector particle carries a total spin angular momentum (J = 1). As a result, only the Right-Left (RL) and Left-Right (LR) configurations contribute to this process [8]. An example of an annihilation process is the production of the Higgs boson in association with a neutral Z boson (Higgsstrahlung) ~\cite{lab8}. To study these processes effectively, choosing an appropriate combination of polarizations is essential for suppressing background radiation and enhancing the signal rate. By increasing the signal-to-background ratio using high luminosity, researchers can perform new searches, even in cases where the rates are expected to be very small.

Alternatively, in the scattering process, the helicities of the incoming beams are independent of each other and directly connected to any resulting particle at the vertex. In this scenario, it is possible for the helicities of $e^{-}$ and $e^{+}$ to be the same, allowing for all four spin configurations ~\cite{lab7}. An example of a scattering process is vector boson fusion, which contributes to Higgs production ~\cite{lab8}. The ability to independently set the polarization of incoming particles is a property that facilitates the exploration of new particle characteristics, such as chiral couplings and quantum numbers, based on certain level assumptions.
Equation~\ref{eq:1} can also be written in terms of the effective polarization

\begin{equation}
\sigma_{P_{e^{-}}, P_{e^{+}}} = ( 1 - P_{e^{+}} P_{e^{-}}) \sigma_{0} [ 1 - P_{eff} A_{LR} ] 
\label{eq:2}
\end{equation}

where $\sigma_{0}$ indicates the unpolarized cross-section, $P_{eff}$ is effective polarization and $A_{LR}$ is left right asymmetry~\cite{lab7}. $\sigma_{0}$ can be written in as:

\begin{equation}
\sigma_{0} = \frac{\sigma_{RL} + \sigma_{LR} }{4} 
\label{eq:3}
\end{equation}

The effective polarization is represented as:

\begin{equation}
P_{eff} = \frac{P_{e^+} - P_{e^-}}{1 - P_{e^{-}} P_{e^{+}}} 
\label{eq:4}
\end{equation}

When the polarized beams are used two enhancement factors in production cross-section are $ ( 1 - P_{e^{-}} P_{e^{+}}) $ and $( 1 - P_{eff} A_{LR})$, while on the other hand when unpolarized $e^+$ beams are used the enhancement factor reduces and cross-section is represented as:

\begin{equation}
\sigma_{i} = \sigma_{0} (1 + A_{LR} P_{e^-} )
\label{eq:5}
\end{equation}

Since there is a significant dependency of LR asymmetry on the polarization and the experimental uncertainty is calculated by the accuracy of polarization. In the case of partially polarized beams, one can easily extract the $\sigma_{0}$ by using eq.~\ref{eq:1}. 

\begin{equation}
\sigma_{0} = \frac{ \sigma_{\mp} + \sigma_{\pm} }{ 2 ( 1 + |P_{e^{+}} || P_{e^{-}}| )}  
\label{eq:6}
\end{equation}

here, $|P_{e^{+}}|$ and $|P_{e^{-}}|$ are representing the absolute value of the polarization.

\begin{equation}
\sigma_{\mp} = \frac{1}{4} \{ [ ( 1 + |P_{e^{+}} || P_{e^{-}}| )( \sigma_{LR} + \sigma_{RL} ) +  ( |P_{e^{+}} | + | P_{e^{-}}| )( \sigma_{LR} - \sigma_{RL} ) \} 
\label{eq:7}
\end{equation}
\begin{equation}
\sigma_{\pm} = \frac{1}{4} \{[ ( 1 + |P_{e^{+}} || P_{e^{-}}| )( \sigma_{LR} + \sigma_{RL} ) -  ( |P_{e^{+}} | + | P_{e^{-}}| )( \sigma_{LR} - \sigma_{RL} )\}
\label{eq:8}
\end{equation}
 
 \section{Dominant Processes at ILC and CLIC}
 For the single Higgs production at linear collider the dominant processes are~\cite{lab9};
 \begin{enumerate}
\item Higgs-strahlung process ($e^- e^+ \rightarrow Zh$)
\item WW-fusion process ($e^- e^+ \rightarrow h\nu_{e} \bar{\nu_e}$)
\item ZZ-fusion process ($e^- e^+ \rightarrow he^- e^+$)
\end{enumerate}

\begin{figure}[ht]
 \centering
   \includegraphics[scale=0.4]{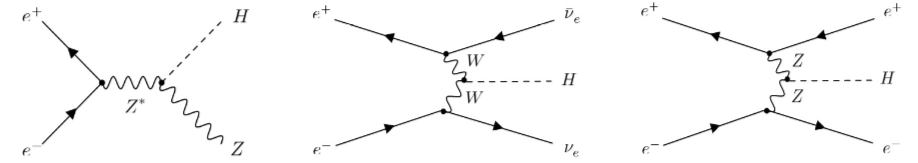}
    \caption{ Feynman diagrams for Higgs-strahlung, $WW$-fusion and $ZZ$-fusion process respectively. }
    \label{fig:1}
\end{figure}

Feynman diagrams for these processes are shown in Fig \ref{fig:1}.

\subsection{The Production Cross-Section of Dominant Processes}
\begin{figure}[ht]
 \centering
   \includegraphics[scale=0.85]{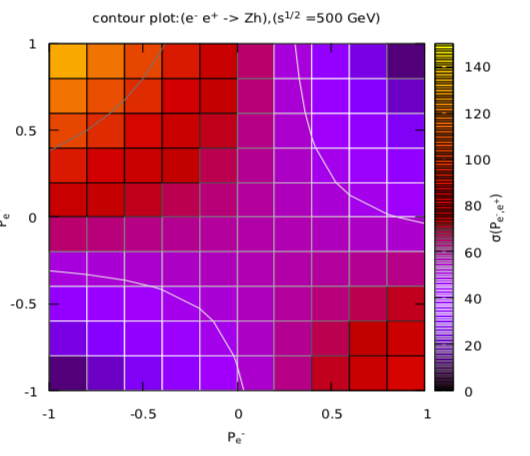}
    \caption{ Contour plot for ($e^- e^+ \rightarrow Zh$) for different possible polarization combinations and corresponding cross-section values. Polarization of electron and positron is represented along $x$-axis and $y$-axis respectively. }
    \label{fig:2}
\end{figure}

Annihilation processes play a leading role at $\sqrt{s}$ = 250 GeV. Higgs self-coupling and its properties can easily be studied through Higgs-strahlung process taking place at high energy collider (ILC). The cross-section measurement of Higgs infers absolute coupling $g(HZZ)$, which serves for the measurement of Higgs width and coupling constant~\cite{lab10}. Now if the effect of beam polarization is observed on the production cross-section of this process, it is configured that different polarization combinations will affect this parameter differently. Therefore, in order to observe this effect a contour plot is drawn for this process at 500 GeV as shown in Fig.~\ref{fig:2}.\\

\begin{figure}[ht]
 \centering
   \includegraphics[scale=0.7]{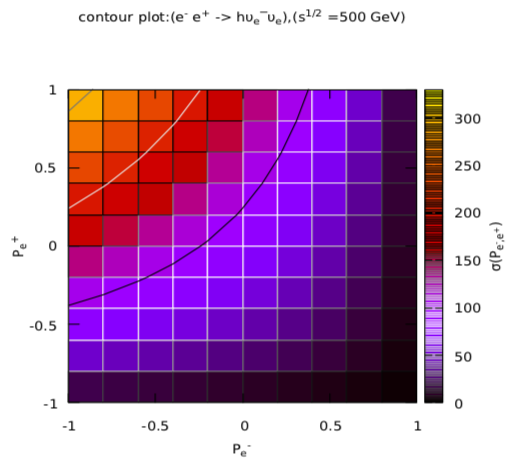}
    \caption{ Contour plot for ($e^- e^+ \rightarrow h\nu_{e} \bar{\nu_e}$) (Left) for ($e^- e^+ \rightarrow he^-  e^+$) (Right), for different possible polarization combinations and corresponding cross-section values. Polarization of electron and positron is represented along $x$-axis and $y$-axis respectively.}
    \label{fig:3}
\end{figure}

From Fig.~\ref{fig:2} it has been observed that for both unpolarized beams, the cross-section is 60.28 $fb$ while maximum cross-section of this process is 144.75 $fb$, when totally polarized left and right-handed electron and positron beams are used, respectively. This value reduced to minimum (0 $fb$), when both the beams are totally left-handed polarized or both are totally right-handed polarized. For other polarization combinations values of cross-section are in between. Similarly, the contour plots for other two processes are shown in Fig.~\ref{fig:3}.
 
\section{Analyzing Higgs using Polarized beams}
\begin{figure}[ht!]
  \centering  
  \includegraphics[scale=0.55]{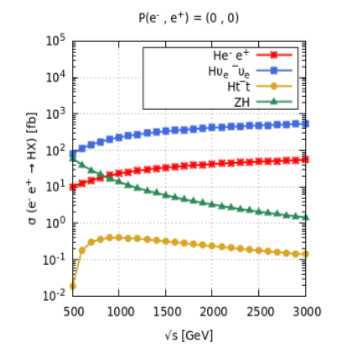}
  \includegraphics[scale=0.55]{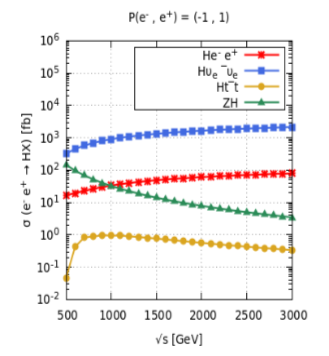}
  \caption{Production cross-section versus CM energy, for dominant processes,using unpolarized beams (left), and fully polarized $e^- e^+$ beams (right) at linear collider.}
  \label{fig:4}
\end{figure}

The properties of Higgs using polarized beams are investigated at $\sqrt{s}$ = 500 GeV. Figure~\ref{fig:4} shows the cross-sections of these processes as a function of centre-of-mass (CM) energy for unpolarized beams. It can be seen from Fig~\ref{fig:4} (left) that at $\sqrt{s}$ = 500 GeV, the two processes have comparable values of cross-section. The cross-section value of $WW$-fusion process increases eventually above this value of CM energy. While on the other hand, the cross-section value of $ZZ$-fusion process is smaller than that of both above mentioned processes at $\sqrt{s}$ = 500 GeV, but value increases in comparison of Higgs-strahlung process above $\sqrt{s}$ = 1000 GeV. Fourth process indicates top Yukawa couplings which is least dominant process. But using fully polarized $e^- e^+$ beams, enhances the value of production cross-section of each dominant process by certain amount as shown in Fig~\ref{fig:4} (right).

\begin{figure}[ht!]
 \centering
   \includegraphics[scale=0.5]{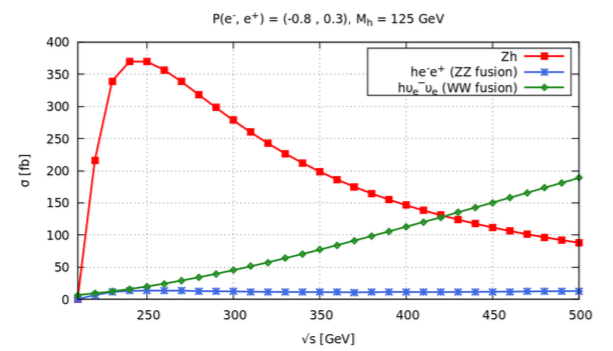}
    \caption{ Production cross-section as a function of CM energy, for three dominant single Higgs production processes.}
    \label{fig:5}
\end{figure}

Higgs reactions can be selected and signal-background mixture can be changed by experimentalists by controlling the polarization of $e^- e^+$ beams. Figure~\ref{fig:5} shows the production cross-section for three dominant single Higgs production processes at $\sqrt{s}$ = 500 GeV. It can be seen that the $WW$-fusion process dominates the Higgs-strahlung process at $\sqrt{s}$ = 500 GeV by using the polarization combination: P($e^+$ , $e^-$) = ( “$0.3"$, “$-0.8"$)~\cite{lab11}.

\subsection{Production processes separation}
\begin{figure}[ht]
 \centering
   \includegraphics[scale=0.6]{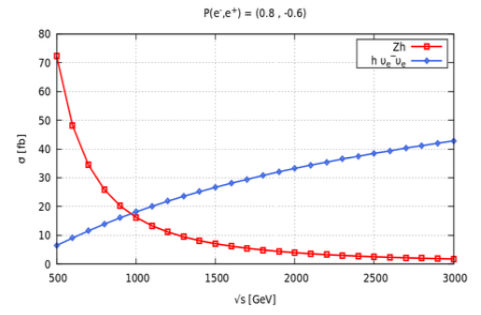}
    \caption{ Production cross-section versus CM energy, for two processes with polarization combination; P($e^+$ , $e^-$) = 
    (  $-0.6$, $0.8$).}
    \label{fig:6}
\end{figure}

At $\sqrt{s}$ = 500 GeV, the Higgstrahlung and $WW$-fusion process have comparable cross-section, but when right-handed (left-handed )polarized beam of electron (positron) is used, it enhances $Zh$ contribution w.r.t $WW$ scattering signal as shown in Fig.~\ref{fig:6}. It also suppresses the background of $WW$- scattering production process which is quite dominant for unpolarized beams~\cite{lab12}. 

\subsection{Top-Higgs Yukawa Couplings}
\begin{figure}[ht!]
 \centering
   \includegraphics[scale=0.45]{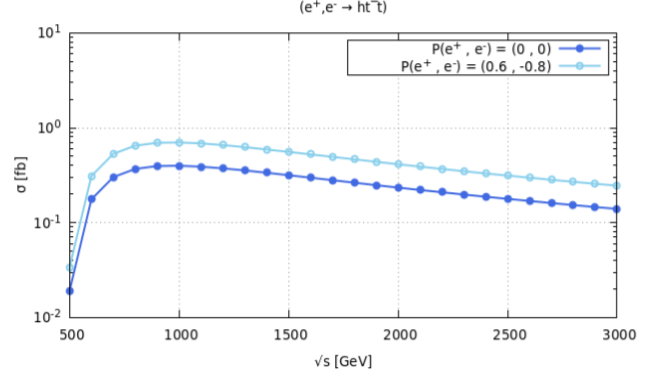}
    \caption{ Production cross-section as a function of CM energy for process $e^{+} e^{-} \rightarrow t \bar{t} H$ }
    \label{fig:8}
\end{figure}

Out of all fermions, there exist greater chances of Yukawa coupling of top quark to Higgs boson due to its mass. In electroweak (EW) symmetry breaking mechanism and generation of mass, it has a major role therefore this coupling needs to be measured precisely. Introducing right-handedly polarized positron beam and left-handedly polarized electron beam; P($e^+$ , $e^-$) = ($0.6$, $-0.8$) , enhances its cross-section value at $\sqrt{s}$ = 500 GeV, $\sigma_{t\bar{t}H}$ = 0.033 $fb$. So its cross-section enhances by a factor 2, which results in precision improvement of $g_{t\bar{t}H}$ of 45\%, and is 24\% for unpolarized beams~\cite{lab12}. The production cross-section as a function of CM energy for the process, $e^{-} e^{+} \rightarrow t \bar{t} H$ is shown in Fig.~\ref{fig:8}.

\section{Additional Parameters}

Additionally, the following three parameters; effective luminosity, effective polarization and left-right asymmetry, mentioned in cross-section formula of processes while utilizing polarized beams are examined. 

\subsection{Effective Polarization}
\begin{figure}[ht]
 \centering
   \includegraphics[scale=0.65]{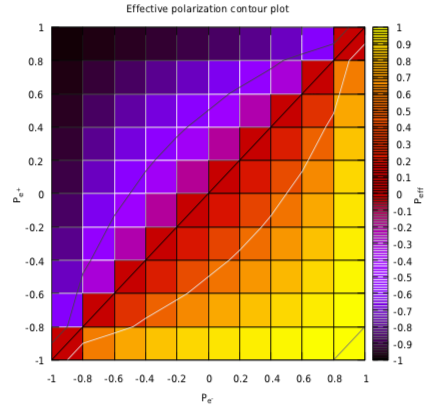}
    \caption{ Plot showing effective polarization values for different polarization combinations.}
    \label{fig:9}
\end{figure}

When both electrons, as well as positrons, are polarized, an effective polarization can be significantly enhanced as given by \ref{eq:4}, the effective polarization value for several polarization combinations can be calculated and then can be observed and shown in Fig.~\ref{fig:9}. In order to achieve the maximum value of effective polarization for a specific polarization combination, only one of the beams must be left-handed polarized, either electron beam or positron beam, not both, as indicated in table~\ref{table:1}.

\begin{table}[h!]
\begin{center}
 \begin{tabular}{|c|c|c|}
\hline
\textbf{ $P_{e^-}$ } & \textbf{ $P_{e^+}$ } & \textbf{ $P_{eff}$ }  \\
\hline
10 $\%$ & 60 $\%$ & 53 $\%$  \\
\hline
-10 $\%$ & 60 $\%$ & 66 $\%$  \\
\hline
10 $\%$ & -60 $\%$ & -66 $\%$  \\
\hline
\end{tabular}%
\caption{The values of effective polarization for different polarization combinations (three particular pairs of values for P).}
\label{table:1}
\end{center}
\end{table}

\begin{figure}[ht]
 \centering
   \includegraphics[scale=0.45]{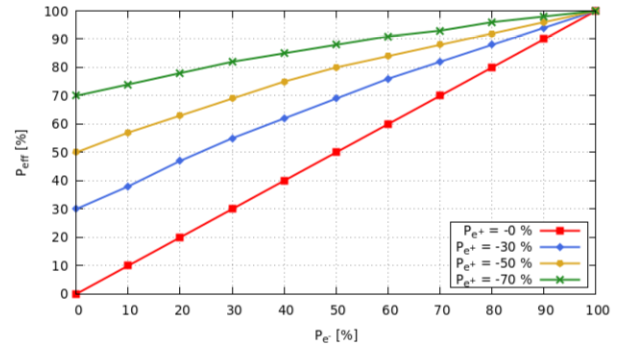}
    \caption{ The Effective polarization as a function of electron beam polarization, while using 0\% or -30\% (left-handed) positron beam polarization.}
    \label{fig:10}
\end{figure}

As when positron beam is polarized, the cross-section value enhances. Similarly to bring enhancement in the effective polarization, the positron beam must be polarized. As can be seen in Fig.~\ref{fig:10}, for $P(e^{+}, e^{-}) = (0\%, 10\%)$, the effective polarization is 10\%, but for $P(e^+, e^-) = (-30\%, 10\%)$, the effective polarization is 38\%.

\subsection{Left-Right Asymmetry}
The asymmetries are special experimental observables estimated by the ratio of observed quantities and the major uncertainties are canceled out. Due to this, the additional independent information on the interaction of particles is provided by asymmetries which are also mentioned in the polarized cross-section formula therefore it is important to measure it precisely with the minimum possible error.

\begin{equation}
A_{LR} = \frac{ \sigma_{LR} - \sigma_{RL} }{ \sigma_{LR} + \sigma_{RL} } 
\label{eq:9}
\end{equation}

for partially polarized beams we can use eq.~\ref{eq:10}

\begin{equation}
A_{LR} = \frac{ \sigma_{\mp} - \sigma_{\pm} }{ \sigma_{\mp} + \sigma_{\pm} } 
\label{eq:10}
\end{equation}

\begin{figure}[ht]
 \centering
   \includegraphics[scale=0.65]{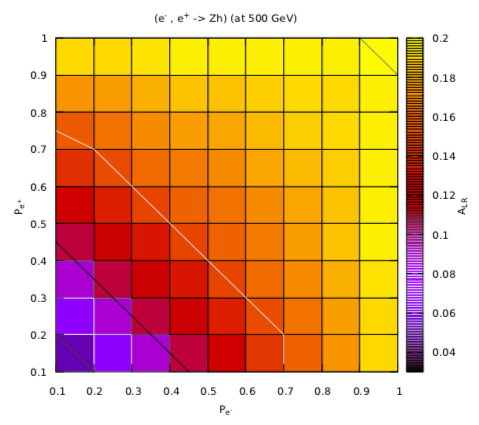}
    \caption{ Plot showing values of $A_{LR}$ for annihilation process at $\sqrt{s}$ = 500 GeV for different polarization combinations.}
    \label{fig:11}
\end{figure}

In charged weak interaction ($t$-channel $W$ or $Z$ exchange), only left-handed fermions and right-handed anti-fermions are involved or take part, therefore, ($\sigma_{LR}$ = 0) for this case and $\sigma_{LR}$ is non-zero or in simple words $A_{LR}$ = 1~\cite{lab8}. The $\sigma_{\pm}$ and $\sigma_{\mp}$ for annihilation process can be calculated by using \ref{eq:5} and \ref{eq:6} respectively, then using in \ref{eq:10}, gives the value of $A_{LR}$ for particular polarization combination. The $A_{LR}$ for several combinations is shown in Fig.~\ref {fig:11}.

\begin{figure}[ht]
 \centering
   \includegraphics[scale=0.65]{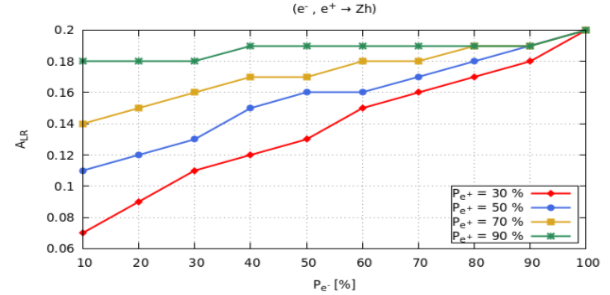}
    \caption{ Left-right asymmetry ($A_{LR}$) electron beam polarization as a function of positron beam polarization.}
    \label{fig:12}
\end{figure} 
In order to check how $A_{LR}$ get affected by the use of polarized positron beam, the Fig.~\ref{fig:12} is plotted between $A_{LR}$ as a function of polarization of electron beam. It has been observed that value of $A_{LR}$ increases with the increase in positron beam polarization. Now the left-right asymmetry formula for the process involving partially polarized beams [$P_{eff} <$ 1] is given;

\begin{equation}
A_{LR} = \frac{1}{P_{eff}} \frac{ \sigma_{\mp} - \sigma_{\pm} }{ \sigma_{\mp} + \sigma_{\pm} } 
\label{eq:11}
\end{equation}

\begin{equation}
A_{LR}^{obs} = \frac{ \sigma_{\mp} - \sigma_{\pm} }{ \sigma_{\mp} + \sigma_{\pm} } 
\label{eq:12}
\end{equation}

where, $A_{LR}^{obs}$ is measured value of left-right asymmetry~\cite{lab7}.
Uncertainty in the measurement of polarization brings about error in left-right asymmetry measurement as,

\begin{equation}
\frac{|\Delta A_{LR}|}{|A_{LR}|} = \frac{|\Delta P_{eff}|}{|P_{eff}|} 
\label{eq:13}
\end{equation}

For the two beams with equal relative precision is given as,

\begin{equation}
x \equiv \frac{\Delta P_{e^-}}{P_{e^-}} = \frac{\Delta P_{e^+}}{P_{e^+}}
\label{eq:14}
\end{equation}

While considering fully correlated relative errors on polarized beams due to depolarization effects, so polarization contribution to $A_{LR}$ uncertainty is as given~\cite{lab7};

\begin{equation}
\frac{\Delta P_{eff}}{P_{eff}} = \frac{1 - P_{e^-} P_{e^+} }{ 1 + P_{e^-} P_{e^+} }\chi
\label{eq:15}
\end{equation}

\begin{figure}[ht]
 \centering
   \includegraphics[scale=0.65]{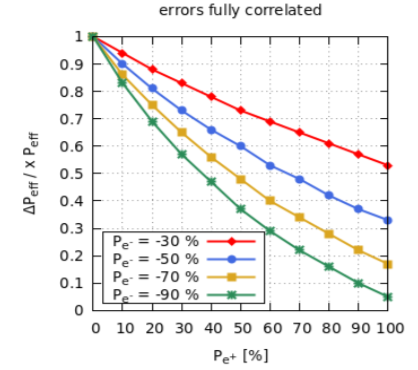}
    \caption{ The dependence of effective polarization uncertainty on beam polarization, normalized to relative precision $x$ for fully correlated errors on beams.}
    \label{fig:13}
\end{figure}

The uncertainty in the effective polarization measurement as a function of positron beam polarization is shown in Fig.~\ref{fig:13}. The different polarization values of electron beam are used and it is clear that positron polarization reduces the contribution of polarization to uncertainty in $P_{eff}$, and hence to error in $A_{LR}$.

\subsection{Effective Luminosity}
Another parameter, effective luminosity, is studied by taking the ratio $L_{eff}/L$, into account, which essentially indicates the percentage of particles that are interacting and is shown in  eq.~\ref{eq:16}.
\begin{equation}
\frac{\ L_{eff}}{L} = \frac{1}{2} (1 - P_{e^+} P_{e^-} ) 
\label{eq:16}
\end{equation}

By considering this, eq~\ref{eq:2} can be rewritten as,

\begin{equation}
\sigma_{P_{e^- , e^+}} = 2 \sigma_{0} (L_{eff}/L) [1 - P_{eff} A_{LR}] 
\label{eq:17}
\end{equation}

\begin{table}[h!]
\begin{center}
 \begin{tabular}{|c|c|c|}
\hline
\textbf{ Polarization value of $e^-$ beam } & \textbf{ Polarization value of $e^+$ beam } & \textbf{ $L_{eff}/L$ }  \\
\hline
$P_{e^-}$ = 0 $\%$ & $P_{e^+}$ = 0 $\%$ & $0.50$  \\
\hline
$P_{e^-}$ = -70 $\%$ & $P_{e^+}$ = 0 $\%$ & $0.50$  \\
\hline
$P_{e^-}$ = -70 $\%$ & $P_{e^+}$ = 60 $\%$ & $0.71$  \\
\hline
\end{tabular}%
\caption{The values of effective polarization for different polarization combinations.}
\label{table:2}
\end{center}
\end{table}

From table~\ref{table:2}, it is indicated that enhancement in the value of fraction of interacting particles can only be observed if both the beams are polarized while the polarization of single beam has no effect. The values of $L_{eff}/L$  are represented in Fig. 13 as a function of electron and positron beams polarization.
\begin{figure}[ht]
 \centering
   \includegraphics[scale=0.60]{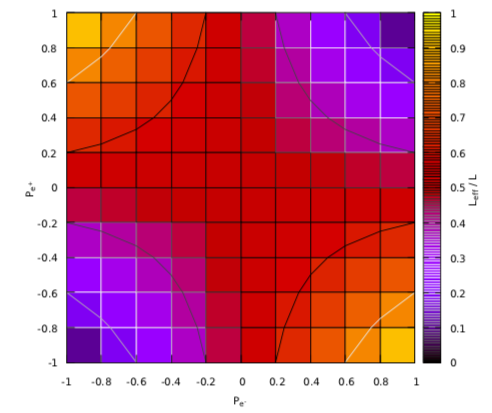}
    \caption{The values of $L_{eff}/L$ as a function of electron and positron beams polarization combinations.}
    \label{fig:14}
\end{figure} 
From Fig.~\ref{fig:15} it is clear that value of $L_{eff}/L$ remains same (0.5), when one of two beam is unpolarized. In order to have greater value of this fraction than 0.5, one of the beam must have to be left-handed polarized. If both the beams are right-handed polarized, then lower value than 0.5 is observed.
\begin{figure}[ht!]
  \centering  
  \includegraphics[scale=0.46]{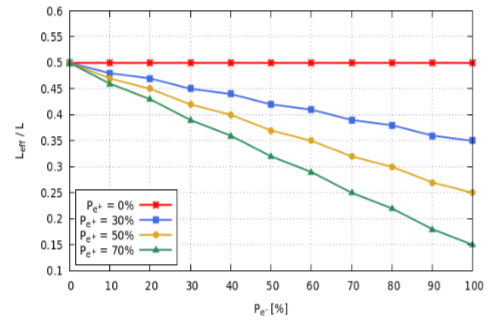}
  \includegraphics[scale=0.48]{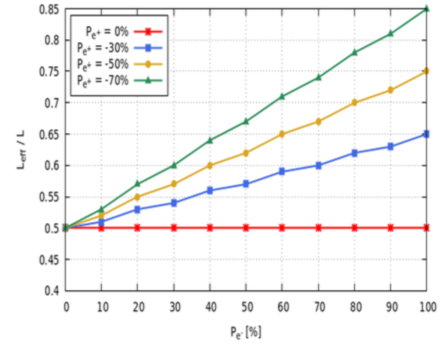}
    \caption{$L_{eff}/L$ as a function of the polarization of electron beam for different values of right-handed polarized positron beam (left) and left-handed polarized positron beam (right.}
  \label{fig:15}
\end{figure}
 \clearpage
 \section{Conclusion} 
The polarity of beams offers a precise instrument for SM testing as well as for the diagnosis of novel physics. The potential of physics can be increased by using polarized beams at linear collider and the structure of this underlying physics could be unraveled more precisely. The physics beyond the SM can be unearthed by the utilization of such polarized beams. Ever since by employing such beams, more observables can be investigated like effective luminosity, actual polarization and left-right asymmetry, as discussed in this current study. The precise measurements of these observables help to measure the production cross-section of a particular process more precisely. 
In the case of top-Higgs Yukawa coupling, which plays a significant part in EW symmetry breaking and mass creation, the utilization of such polarized beams results in precision. \\
New physics phenomena can be discriminated like in the production processes selection through which decay modes of Higgs and its production rate can be measured. Precision and high accuracy is only possible for the different observables if we use equally the polarized positron and polarized electron beams. The polarization sources are being established at CLIC and ILC, but it is still a challenging task to establish them at FCC, this work is an understudy.\\

\section{Acknowledgment}
We gladly encouraged the Simons Foundation and Cornell University member institutions for their assistance. The present submitted form of manuscript is available on arXiv pre-prints home page https://arxiv.org/pdf/2202.13567.pdf with arXiv ID:2202.13567.

\section{Statements and Declarations}
\textbf{Funding} \\
The authors declare that no funds, grants, or other support were received during the preparation of this manuscript.\\
\textbf{Competing Interests}\\
The authors have no relevant financial or non-financial interests to disclose.\\
\textbf{Author Contributions}\\
Mashal Shakar being master student under my supervision worked on beam simulation and results extraction. Ijaz Ahmed wrote the manuscript partly contributed by M.U.Ashraf and Mashal Shakar. M.U. Ashraf and Jamil Muhammad both thoroughly reviewed the entire manuscript, performed proof reading and grammatical checks, removal of formatting issues, corrections in references etc.\\

\textbf{Availability of data and materials}\\
Data sharing not applicable to this article as no datasets were generated or analysed during the current study.\\

\end{document}